\documentstyle[12pt,aasms]{article}
\input{ref.macro_apj}

\def\gs{\mathrel{\raise0.3ex\hbox{$\scriptstyle >$}\kern-0.70em %
\lower0.71ex\hbox{{$\scriptstyle \sim$}}}}
\def\ls{\mathrel{\raise0.3ex\hbox{$\scriptstyle <$}\kern-0.70em 
\lower0.71ex\hbox{{$\scriptstyle \sim$}}}}
\def\et{\hbox{\it et al.}$\,$}
\def\go{
\mathrel{\raise.3ex\hbox{$>$}\mkern-14mu\lower0.6ex\hbox{$\sim$}}
}
\def\lo{
\mathrel{\raise.3ex\hbox{$<$}\mkern-14mu\lower0.6ex\hbox{$\sim$}}
}
\def\w{\omega(\theta)}
\def\A{A_\omega}

\lefthead{Hudon and Lilly}
\righthead{Clustering of Galaxies to $z$=0.6}

\begin{document}

\title{The Clustering of Faint Galaxies and the Evolution of $\xi$(r)} 

\author{J. Daniel Hudon\altaffilmark{1} and Simon J. Lilly\altaffilmark{1}}
\affil{Department of Astronomy, University of Toronto, Toronto, Canada }

\altaffiltext{1} {Visiting Astronomer, Canada-France-Hawaii Telescope, which
is operated by the National Research Council of Canada, 
the Centre National de la Recherche Scientifique of France and the University of Hawaii}
\begin{abstract}
The two-point angular correlation function, $\w$, is constructed from
a catalog of
13,000 objects in 24 fields distributed over an area of 4 deg$^2$
and complete to a limit of $R = 23.5$. 
The amplitude and slope of
our correlation function on arcminute scales are in broad agreement with 
recent CCD results in the literature and decreases with depth.
No evidence is found for a flattening of the slope of the correlation function
away from $\delta \sim 0.8$.
Using the redshift distribution from
the recent I-band selected Canada-France Redshift Survey, the observed $w(\theta)$ implies 
a value of the clustering length $r_0 = 1.86 \pm 0.43 \h$ Mpc ($q_0 = 0.5$)
at $z = 0.48$.
This is generally consistent with the possible rates of clustering 
evolution expected for optically selected galaxies. We finally discuss
the implications of our results for the nature of the faint
galaxy population.

\end{abstract}


\section{Introduction}

Understanding
the strength and evolution of field galaxy clustering is a key ingredient
to synthesizing several areas of current extragalactic research. The evolution 
of the galaxy correlation function will likely reflect both the evolution of
large scale structure in the Universe and the effects of evolution on the galaxy population.
The 
large-scale distribution of galaxies is a product of the initial density
fluctuations present in the early Universe and the growth of these fluctuations
is strongly $\Omega$ dependent. 
The strength of clustering provides clues as to the nature
of the galaxies themselves, for example, whether they are dwarf galaxies,
normal galaxies or low-mass haloes in the process of forming galaxies.
Finally,
as merging may be one of the drivers of galaxy evolution, there may be a direct
link between the evolution of clustering and the evolution of galaxies.

As new data on galaxies at cosmological distances is acquired, we are able
for the first time to examine the correlation function at redshifts
that are sufficiently large that the evolution in the correlation function
should be unambiguously detectable.

An observationally convenient statistic to quantify galaxy clustering is
the two-point correlation function, $\xi(r)$, and
its two-dimensional projection onto the surface of the sky, $\w$, both 
of which measure the strength of clustering in comparison to a random
population. 
From bright surveys, $\w$ has been found to obey a power-law of the form
$\w = \A\theta^{-\delta}$ where $\delta = 0.8$ (Peebles 1980). 
More recently, Maddox \et (1990) have found $\delta = 0.66$ for galaxies
between $17.5 \leq b_J \leq 20.5$ in the APM survey.
Regardless of any cosmic evolution in the clustering, 
the amplitude of $\w$ will decrease with increasing magnitude (i.e. depth) 
because the increased path length means that an increasing fraction of 
pairs at a given angular separation are unrelated chance projections. 

In this paper, the angular correlation function is determined for a sample 
of galaxies selected in the
R-band, $19 < R < 23.5$. At the median redshift of this sample ($z \sim 0.56$),
the R-band samples the rest-frame B-band light of the galaxy 
population and the selection of this sample is therefore well-matched to that
of local samples of optically-selected galaxies.
Recent papers on the correlation function in the R-band include those of 
Infante and Pritchet (1995)
who used photographic plates to measure correlations of 39000 galaxies 
over an area of 2 deg$^{2}$ and to a depth of F = 23.0. They found 
evidence for a flattening of the slope of the power-law with increased depth
between $F=21$ and $F=23$. Also, clustering amplitudes were closer to
no-evolution model predictions in the red filter (F) sample than in the
blue (J) filter sample.
Couch \et (1993) examined $\w$ for more than 116,000 galaxies over an
area of 3.6 deg$^{2}$ in a hybrid ``VR'' band to an equivalent depth of
$R \sim 23$ where they found no evidence for changes in the slope with depth.
Roche \et (1993) used 3627 galaxies distributed over 331 arcmin$^{2}$
to measure correlations $\w$ to a depth of $R = 23.5$, obtaining  
reasonable agreement with a $\theta^{-0.8}$ power law. 
Efstathiou \et (1991) studied correlations between $23 < R < 25$ (as well as
$24 < B_{J} < 26$) and found a surprisingly low clustering amplitude,
i.e. their correlations at $\theta \leq 0^{\circ}.06$ showed little or
no signal. 
The deepest survey thus far
is that of Brainerd \et (1995) who used Keck data to $r = 26$. They observed
5700 galaxies on a single 90 arcmin$^{2}$ CCD field and found the familiar
$\theta^{-0.8}$ power law along with a low but 
non-zero value for the correlation amplitude. 
Lower than expected correlation amplitudes can be produced by a broadening
of the $N(z)$ (Efstathiou \et 1991, Roche \et 1993 ) or if the sample
is dominated by dwarf galaxies that are more weakly clustered than
bright galaxies (Efstathiou \et 1991, Brainerd \et 1995)

In this paper we use high quality images taken in sub-arcsecond seeing 
with the 3.6m 
Canada-France-Hawaii Telescope to investigate the angular correlation
function from arcsecond to arcminute scales to a depth of $R = 23.5$.
In an important development, the results from the deep 
I-selected, Canada-France
Redshift Survey (see e.g. Lilly \et 1995a, Le F\`evre \et 1995, 
Crampton \et 1995) 
are used to determine the $N(z)$ for our R-band selected sample. 
This information enables us to estimate the true three-dimensional 
correlation function $\xi (r)$ at a fiducial redshift of our 
sample, $z \sim 0.48$ (see Section 4.2).

The layout of this paper is as follows. In Section 2, the data reduction and image analysis which 
generated the photometric galaxy 
catalogs is described. In Section 3, our methods of correlation estimation
and bias correction are described and our estimate of the $\w$ on 
angular scales up to a few arcminutes presented. These results are compared 
with those of others in section 4, and using our new knowledge of $N(z)$, 
an estimate for the 
$\xi (r)$ at $z = 0.48$ is derived.  The implications of this measurement 
are discussed in 
in Section 5. Throughout the paper, we take $H_{0}$ to be 
100 h kms$^{-1}$Mpc$^{-1}$ and assume 
$q_{0} = 0.5$ unless otherwise indicated. Length scales, such as correlation 
lengths, are quoted as physical (or ``proper''), rather than comoving, lengths.

\section{Data}

Characteristics of the data and its calibration are described below. Further
details may be found in Hudon (1995).

\subsection{Observations and Data Reduction}

The primary dataset for this project consists of R-band CCD images 
obtained with the Faint Object Camera (FOCAM) at the Prime Focus of the 3.6m 
Canada-France-Hawaii Telescope on Aug. 10-12, 1991.
Twenty-four fields randomly distributed across an area of 4 deg$^2$
were observed, with each field covering an area of approximately 50 
arcmin$^2$. 
Assuming a redshift interval $0.3 < z < 0.7$ (see below), the survey
samples a volume of $\sim 10^6 h^3$ Mpc$^3$.
The observations
were made under good seeing conditions ($0.''6 - 0.''9$) using the LICK2
chip with pixel size 0.207 arcsec. 
For reasons unconnected with this project, 
each of the 24 fields was imaged on three occasions over the course of 
three nights for a total 
exposure time of 15 minutes. The separation in time between exposures thus 
varied between
10 minutes and 2 days. Photometric calibration through the run was
carefully monitored with repeated 
observations of three Landolt (1993) standards,
together with a reference star 
that was located near to our fields. 
The color term was found to be essentially zero.
The survey is centered on the SSA-22 field (2215+00) of Lilly, Cowie and 
Gardner
(1991, see also Lilly 1993) which is also one of the five fields of 
the Canada-France Redshift Survey (see e.g. Lilly \et 1995b).

The CCD images were each pre-processed by subtracting a median bias frame and
flat-fielded using first dome flats and then a sky-flat generated from the 
whole data set. 
After this procedure, a low-amplitude fringe
pattern of variable amplitude persisted, caused primarily by an imperfect 
coating on the CCD chip. We 
developed an iterative scheme to effectively separate out and remove the 
pattern from each of the images.

Each image was photometrically calibrated individually, and then the three 
images for each field were coadded using a 3-sigma clipping algorithm. This
was effective in removing essentially all cosmic rays.
The photometric zeropoint for the three observations of each field
was found, from photometry of stellar images in each field, 
to be consistent to better than 2 \% rms.

\subsection{Catalog Generation}
Object detection and photometry on the final co-added images was done using 
FOCAS (Valdes 1982). We used detection criteria of 
$3\sigma$ per pixel and a six pixel minimum contiguous area.
As we believe has been seen before, 
various tests such as generating catalogs from rotated images or after 
the addition of a constant background level produced slightly
different catalogs. Image rotation produces different catalogs because
the image detection algorithm proceeds through an image row by row and thus
the local sky at any object depends on the history of pixels leading up to
that object. However, it was found that merging together the 
catalogs from images 
rotated by $0^\circ, 90^\circ, 180^\circ$ and $270^\circ$ with the
background sky level determined independently from FOCAS produced a more
complete catalog in the sense that fewer objects were missed.
Approximately 1000 objects were found on each image down to our magnitude
limit of 24.2.  The catalogs were examined on an interactive 
display using the Picture Processing Package (Yee 1991) and spurious 
objects removed. 
These were typically due to haloes around bright stars or multiple
identifications of faint amorphous objects, and, together, these comprised 
a few per cent of the catalogs. A handful of obvious objects missed by the 
image detection algorithm were also added at this stage. 
Due to a small spatial shift between the three images in each field, we have 
used only objects that are located more than
48 pixels from the edge of the image as the image edges suffer from dead 
pixels, cosmic rays, and greater photometric errors due to the coadding. Using this finding list, 
magnitudes were determined using a fixed aperture of 5$''$. Magnitudes
determined in this way are within 0.2 of ``total'' magnitudes.
The 1$\sigma$ photometric noise on these large apertures is 
equivalent to an R-magnitude of approximately 24.2.

\subsection{Number Counts}

The number counts of deep objects provide a good consistency check 
between our catalogs and others who have examined the counts at similar depths.
Following Couch \et (1993) we have plotted our counts together with the
compilation of Metcalfe \et (1991, their Figure 11) in Figure~1. 
Our counts are seen to be in good agreement with other recent determinations
in both slope and absolute number. The best-fit slope of the solid line
in Figure~1 is 0.36.
Stars dominate
the catalogs at magnitudes brighter than 19. This is seen in the data
(starred symbols) as well as the predicted star counts of Bahcall and
Soneira (1980) (see Section \ref{ssec:sc} below). 

The turnover in our counts in the faintest two bins is due to 
incompleteness. 
For the correlation analysis below, we decided to take a conservative
magnitude cut-off of $R = 23.5$, which is above the incompleteness roll-over 
in all
of the individual fields. The standard deviation of the number of objects 
observed from
one field to the next down to $R=23.5$ is 15\%.  This did not correlate 
with any obvious observational parameter such as the seeing, airmass etc., 
and is most likely to be dominated by large scale galaxy clustering 
(see Section 3.3 below).

\subsection{Stellar Contamination}
\label{ssec:sc}
Foreground stars are randomly distributed on deep sky images and have the 
effect of diluting the amplitude of $\w$ relative to that
which would be measured for the galaxies alone.  It can easily be 
proved that the stellar-corrected
correlation function is equal to the observed correlation function divided
by the square of the galaxy fraction. 
For reasons unconnected with this project, the survey fields are at relatively 
low Galactic latitude $b_{II} = +46^\circ$, so stellar comtamination is 
relatively important.
This problem may be addressed either by removing stars from the sample or
by correcting the $\w$ determined from a composite sample that includes the 
stars. Although at this seeing and S/N the vast majority of galaxies are 
distinguishable
morphologically from stars (see Crampton \et 1995), a danger in attempting
to remove stars from the sample is that the criteria that separate stars from
galaxies can vary with magnitude and, especially, from one 
field to the next
due to variations in the point spread function. 
There is also the possibility of systematic effects if the most compact
galaxies have different clustering properties.
Accordingly, the second approach was adopted in this investigation.

Estimation of the stellar contamination proceeded as follows. 
First, at bright magnitudes,
where there are relatively few objects,
a star-galaxy separator was applied which
compares the intensity of the peak pixel for each object to the average
intensity of the remaining pixels in a 3 arcsecond aperture. On our
images, the distribution of stars and galaxies in the plane overlaps at a 
magnitude of $R \gta 22.0$,
as shown in Figure~2 for four of the 24 different fields.
Beyond this limit, we used the galaxy model of Bahcall and Soneira (1980)
to estimate the faint star counts. 
Finally, we checked our star counts with those
derived from the I-band selected Canada-France Redshift Survey (which obtained 
spectra of all objects regardless of morphology)
in this same region of the sky (Lilly \et
1995b) and found good agreement.
Our estimates for the 
galaxy fraction, $f_g$, as a function of depth are presented in Table 1.

\subsection{N(z) Information}

The redshift distribution is a critical element in interpreting the observed
amplitude of the projected 2-d correlation function $w(\theta)$. 
The Canada-France Redshift Survey (CFRS, see e.g Lilly \et 1995a, Le F\`evre
\et 1995, Crampton \et 1995) is a new
spectroscopic survey of over 1000 red-selected objects with isophotal
$17.5 < I_{AB} < 22.5$, comparable in depth to the galaxy sample considered
here. The overall success rate in securing redshift measurements for the
galaxies was 80\%. However, as discussed by Crampton \et (1995), the redshifts of half
of the unidentified objects are known statistically, and
it is likely, based on their colors, that the remaining 10\% of galaxies do
not have a very dissimilar redshift distribution. The fraction of the sample 
for which redshift information is not available is therefore less than 10\%.

Since
($V-I$) colours are available for all objects,
and ($I-K$) and ($B-I$) for most of them, it is relatively straightforward
to produce a predicted N(z) for an R-band selected sample. This procedure, 
described by Lilly \et (1995c), uses the $1/V_{\rm max}$ formalism. For
each galaxy, the absolute magnitude and spectral energy distribution
is defined, and thus the
comoving volume throughout which it would have been 
within the apparent magnitude limits of the original
I-band selected sample ($17.5 \leq I_{AB} \leq 22.5$) is calculated:

\begin{equation}
V_{\rm max} = \left( {c \over H_0} \right)^3 \int_{z_{17.5}}^{z_{22.5}}
\frac{Z_q^2(z)}{(1+z)(1+2q_0z)^\frac{1}{2}} d\Omega dz,
\end{equation}
$d\Omega$ is the effective solid angle of the survey, 112 arcmin$^2$
(see Le F\`evre \et 1995)
and $$Z_q(z)= \frac{q_0z + (q_0-1)[\sqrt{1+2q_0z}-1]}{q_0^2(1+z)}. $$ 
The spectral energy distribution is determined by interpolating
the spectral energy distributions given by Coleman \et (1980) on the basis of the 
observed $(V-I)_{\rm AB}$ colours, available
for all objects.

Using the same absolute magnitude and spectral energy distribution, the apparent R magnitude 
as a function of redshift of each galaxy is calculated and the $(R,z)$ plane is
populated with

\begin{equation}
dn = \frac{dV}{dz}\frac{1}{V_{\rm max}}.
\end{equation}
Integration over R then gives the N(z) distribution
for a given R-selected sample.
The procedure should account fully for 
the different 
weighting of galaxy types in the R-band sample, and for the 
effects of K-corrections and volume elements as the redshift increases. 
The procedure does not, however, attempt to account for any evolutionary 
effects (i.e. it is assumed that the population does not evolve at higher
or lower redshifts than the galaxies observed, 
but for samples well-matched in redshift (i.e. with similar $<z>$), these 
effects should be small.

In constructing the R-band N(z) we used photometrically estimated redshifts
for all the CFRS galaxies for which reliable spectroscopic redshifts were
not available (the ``best estimate'' sample of Lilly
\et 1995b).
The redshift distribution derived in this way for our photometric sample 
$19.0 < R < 23.5$ 
and several other magnitude ranges is shown in Figure ~3.
We note in passing that although the predicted $N(m,z)$ must reproduce the 
number-magnitude counts at the depth of the original survey, it does not do 
so at substantially fainter magnitudes. This is presumably due to continuing 
evolution of the
luminosity function at higher redshifts (see Lilly \et 1995c for a discussion).

The effect of placing galaxies with unknown redshifts at some arbitrary 
redshift
is to dilute the projected correlation function, leading to an underestimate
of the true 3-d clustering in the sample. The maximum effect would be observed if the
unidentified objects were, like the foreground stars, intrinsically unclustered (perhaps
because they lay at very high redshift). A 10\% contamination by unclustered 
galaxies, the maximum that we could conceive,
would have a 20\% effect on the amplitude of the correlation function and
therefore a 10\% effect on the estimation of $\rnot$.

\section{Calculation of the Correlation Function}

\subsection{The Estimators}
The power of the projected correlation function, $\w$, is its ability to 
deliver
clustering information for a magnitude-limited sample of galaxies simply
by taking images of the sky, without recourse to
time-consuming individual distance measurements. 
In addition, since
$\w$ is not affected by distortion due to the peculiar motions of galaxies,
and photometric surveys probe deeper than spectroscopic surveys,
$\w$ is effectively as useful as $\xi(s)$ itself.
The correlation function $\w$ is defined as the excess probability (above random) of
finding a galaxy in an angular element $\delta\Omega$ at a distance
$\theta$ away from some reference galaxy:

$\delta P = N (1 + w(\theta)) \delta\Omega, $

where N is the mean density of objects on the sky.
Over the last two decades, several
methods have been developed to estimate the above probability. These
are discussed in detail by Sharp (1979), Hewett (1982)
and Infante (1994) amongst others. 
For this investigation, we have used both the counts-in-cells method
in which the form of the estimator is:

\begin{equation}
w(\theta) = \frac{\langle N_{i}N_{j}\rangle}{\langle N_{i}\rangle\langle N_{j}\rangle} - 1            
\label{eq:w1}
\end{equation}
where $N_{i}$ and $N_{j}$ denote the counts in cells i and j
and the brackets
denote averages over pairs of cells separated by $\theta \pm \delta\theta / 2$.

In practice, each image was divided up into cells 32 pixels on a side
(i.e. $6.''4 $ x $ 6.''4$) and $w(\theta)$ was calculated for bins that were 
separated by 0.15 in log$\theta$. The range is $5.''5  \leq \theta \leq 204''$,
the upper limit corresponding to half the size of each individual CCD image. 
The large bins cause some ``digitisation'' effects at the smallest scales
where the effective separation is no longer the nominal separation between the cell centers.
The smallest two bins have been adjusted to smaller effective separations
to reflect this and on these small scales we have also constructed traditional
ring counts which agree well.

\subsection{Errors}

The internal errors in our correlation function can be estimated in three 
different ways:
(1) Modified Poisson errors from the number of pairs at any separation 
(e.g. Peebles 1980, $\S$ 48):
$\delta\w = \sqrt{ \frac{1+\w}{N_{pairs}(\theta)}}$,
(2) by taking the standard error of the 24 averaged correlation functions
(i.e. from the ``local'' method, discussed below), or (3) from the bootstrap 
method (e.g. Ling \et 1986).
Since the bootstrap method is a general methodology for assessing the
accuracy of a given estimator (Efron and Tibshirani 1986), we choose it to 
evaluate our correlation uncertainties.
For the bootstrap method, given the original data set, a pseudo-data set is 
generated by choosing N
data points with replacement from the original set of N data points and the
correlation function is redetermined. This process is repeated a minimum
of 300 times. At each separation, $\theta$, we take the error as the 
standard deviation
of the distribution of points in the pseudo-correlation functions.
We note that our bootstrap errors are approximately 40\% larger
than the error determined from the standard deviation of the correlation
function for individual fields or Poisson errors.

\subsection{Biasses in $\w$}

A key issue in the construction of $\w$ is the estimation of the background 
density of galaxies.
The estimator for $\w$ may be biased for two reasons: (1) the ``integral
constraint'' (Peebles 1980) and (2) possible variations in the surface 
density of galaxies introduced by
spurious observational effects especially from field to field.
The integral constraint arises because the galaxy surface
density is estimated on finite angular scales.  If $\w$ is still positive 
on these scales, the galaxy surface density, and hence the expected number
of pairs, is biased. Thus, the observed $\w$ is reduced from its true 
value and forced to
become negative at some large angle, roughly half the angular extent of the 
sample. 
The integral constraint operates as a scaling factor, $B$ 
(with $B < 1$), introduced into the
estimator to correct for this bias, {\it viz.},
\begin{equation}
w(\theta) = \frac{\langle N_{i}N_{j}\rangle}{B\langle N_{i}\rangle\langle N_{j}\rangle} - 1.            \label{eq:w1b}
\end{equation}

Note that for convenience, we have allowed the integral constraint to 
operate as a multiplicative scaling factor; since $\w << 1$, this is
not significantly different from the usual additive way of accounting
for the integral constraint (Peebles 1980, $\S 32$).
The value of B may be estimated if the true $\w$ is known (e.g. from the
double integral of $\w$ over the whole sky). For a power-law
angular correlation function with a cutoff in correlation power beyond
some angle $\theta_{c}$, B is given by (Pritchet and
Infante 1986),
\begin{equation}
B = \left[ 1 + \frac{2 A_{\omega}}{2 - \delta}\left(\frac{\theta_{c}}{\theta_{o}}\right)^{2}{\theta_{c}}^{-\delta} \right]^{-1},
\end{equation}
where $\theta_{o}$ is the radius of the field being observed, $\A$ is the 
amplitude at $1^{\circ}$ and $\delta$ is the slope.  (For 
$\theta_{c} > \theta_{o}$, $\theta_{c}$ should be replaced with $\theta_{o}$ in
the above formula.)

If there are spurious field-to-field variations (e.g. from errors in the
photometric zeropoint) then the effect is to bias $\w$ to higher 
values (because the 
spurious variations mimic galaxy clustering on large scales). This introduces a
second parameter $B'$ (with $B' >$ 1), given by,
\begin{equation}
B' = 1 + \left(\frac{N_{inst}}{N_{avg}}\right)^{2},
\end{equation}
where $N_{inst}$ is the rms variation in the number of objects due to
spurious instrumental (photometric) error and $N_{avg}$
is the average number of objects per frame.

As noted above, the numbers of galaxies down to a fixed magnitude limit 
varies by 15\% across our sample. The photometric calibration is known to be 
constant to within 2\% rms.
Hence, purely photometric uncertainties would only produce
a $\sim 2\%$ variation in the expected numbers of galaxies (given the slope 
of the number counts) and thus a negligible
$4*10^{-4}$ bias to $\w$. Although we could find no correlation between the 
numbers of galaxies observed
in each field and other observational parameters such as the seeing or 
airmass of observation, we
have no way, a priori, of knowing what fraction of this 15\% variation is 
due to observational effects
and what fraction is due to true galaxy clustering.  

In the light of this, we have taken two approaches to the construction of $\w$.
First, we compute $\w$ in each field individually, i.e. the sum over cells is
carried out over each field separately, and the resulting $\w$ is then 
averaged.
This is effectively estimating the background galaxy density
from each field separately which obviously eliminates any problems due to 
spurious field-to-field variations, but produces a relatively large integral 
constraint bias. We refer to this as the ``local" determination of $\w$. 
Second, the sum over cells is carried out over all cells in the sample, 
effectively setting the background galaxy density to be the average of the 
whole sample, determined over a scale almost 20 times larger. We refer to 
this as our ``global'' $\w$. It has a very small integral constraint bias
but may suffer from spurious field-to-field variations, although, as 
noted above, in our sample these should be
small (at least from photometric variations).

The correlation function for the counts-in-cells method is shown in
Figure~4 for three magnitude samples: $19 \leq R \leq 22.5$,
$19 \leq R \leq 23.0$ and $19 \leq R \leq 23.5$. We have presented both 
the local and global averaging methods along with a power-law,
$\theta^{-0.8}$. The data points are uncorrected for the 
integral constraint, spurious field to field variations or
for the stellar contamination.  As discussed above, 
at larger scales the integral
constraint pushes the local function down while spurious 
field-to-field variations may
push the global function up.   The best fit slopes to the global
$\w$ are $\delta = -0.82$ for $19 < R < 23$ and $\delta = -0.73$
for $19 < R < 23.5$.  The decrease in the last half magnitude is
almost certainly due to the effect of small spurious field-to-field variations
entering towards the bottom of the sample and
introducing a non-zero B$'$ term, thus raising the correlation function at 
the largest scales.
These slopes give little support to the
suggestion that the slope of $\w$ flattens with depth 
(c.f. Infante and Pritchet 1995).

Together, the local and global 
methods must bracket the true (unbiased) correlation function.  To obtain an 
estimate for this
true correlation function for $19 < R < 23.5$, we follow other recent work and
force the slope of the true unbiassed $\w$ to be $\delta = -0.8$ --
the observed slope from local surveys (Peebles 1980) and as
observed in our $19 < R < 23$ sample.
This effectively determines both B and B$'$ empirically (subject, of
course, to the assumption
about the slope).

In the $19 < R < 23.5$ sample, we find B = 0.987 for the locally estimated 
$\w$ and B$'$ = 1.004 for the global estimate.   The value of B agrees 
exactly with the Pritchet and Infante formula
(as it must), while the small value of B$'$ implies that the observed field 
to field variations
of 15\% rms are dominated by true clustering effects. The component
arising from observation effects is (again, assuming the slope of 
$\delta = -0.8$)
about 6\% rms in the faintest sample.  This is about 3 times larger 
than that expected from photometric errors above, but is not too surprising 
and, as noted above, these spurious effects disappear entirely for the samples 
limited at slightly brighter magnitudes, $19 < R < 23.0$,
for which the observed variations are entirely accounted for by the 
expected correlations of galaxies from field to field with $\delta = -0.8$. 

We summarize our results in Table 1.

\section{The correlation function $\w$ on arcminute scales}

\subsection{Comparisons with the results of others}

In order to compare our correlation function with those of other recent 
investigations, 
we have determined the amplitude of the correlation function extrapolated to 
a scale of $\theta = 1^{\circ}$ for a $\delta = -0.8$ slope.
It should be noted, however, that this amplitude is based on the amplitude 
of $\w$ on scales up to 2 arcmin, i.e. on Mpc scales at $z$ = 0.48.

The amplitude at $\theta = 1^{\circ}$ is
compared with other recent determinations, extrapolated with the same 
$\delta=0.8$ slope, in Figure~5,
where the general decline of correlation amplitude with limiting survey depth
is clearly evident.
Data at the bright end is taken from Stevenson \et (1985)
who used 1.2m UKST plates to carry the correlation analyses to $r_F < 20$
(we have neglected to plot their AAT data as they report anomalously low
number-magnitude counts for one of their fields).
Our amplitudes agree well with those of Infante and Pritchet (1995) and 
Roche \et
(1993), but we are approximately two times 
higher than
Couch \et (1993). Fainter than $R = 23.5$, the correlation amplitude appears
to drop off more steeply with the low results of Brainerd \et (1995) and
the single measurement of Efstathiou \et (1991).
All points in the figure have been corrected (either by the original 
authors or by ourselves) for contamination by foreground stars.

\subsection{Inversion to yield $\xi(r)$}

With the completion of the I-band selected Canada-France Redshift Survey
and the definition of $N(z)$ for red-selected faint galaxy samples, 
it is possible to invert the projected $\w$ to yield an estimate for
the three-dimensional $\xi(r)$ at some fiducial redshift.  

Since the spatial correlation function
is observed to be a power law locally (Davis and Peebles 1983) and to have 
essentially the same slope at fainter magnitudes (see above), it is 
convenient to parametrise
the cosmic evolution of clustering by a parameter, $\epsilon$, such that
(Groth and Peebles 1977, Phillipps \et 1978):

\begin{equation}
\xi(r,z) = \left( {r_{0}(0) \over r} \right)^{\gamma}
           \left(1+z \right)^{- \left( 3+\epsilon \right)},
\end{equation}
where $r_0$ and r are measured in physical units, and $\rnot$ is the
correlation length at $z=0$.

The change in physical correlation scale length with redshift, $r_{0}(z)$ 
is thus:
\begin{equation}
r_{0}(z) =  {r_{0}(0) \over (1+z)^{ (3+\epsilon)/\gamma}}.
\end{equation}
The quantity $r_{0}(z)$ is simply the correlation length that would be 
measured, in physical units, by an observer at the epoch in question.  
The value of $r_{0}(z)$
that would be expected at some earlier epoch clearly depends on the 
correlation length of that population at the present epoch, $r_{0}(z=0)$, 
the slope of the correlation function,
$\gamma$, and the value of the evolutionary parameter, $\epsilon$.

The evolutionary parameter $\epsilon$ is interpreted as follows: 
$\epsilon = -1.2$ corresponds to the case where clustering is fixed
in comoving coordinates -- galaxy clusters expand with the Universe, there is 
no
relative motion of galaxies and clustering does not grow with time;
$\epsilon = 0$ is produced by clustering models that have bound units of 
constant physical size
(i.e. in the same way that galaxies are bound objects that do not grow with
the expanding Universe). The clustering grows in this case because the 
background density of galaxies is diluted by the expansion (while the density 
in the clusters is constant) - effectively it is the voids that are growing.
As an example of an evolutionary
model with hierarchical growth, we have examined the mass correlation function
from the Cold Dark Matter simulations of Davis et al (1985). The 
$\Omega = 1$ EdS models have $\epsilon \sim +0.8$, since $\xi$ in comoving 
terms evolves as $(1+z)^{-2}$ as expected from the linear growth
of perturbations.

Once the spatial distribution of galaxies in a sample is determined 
through $dN/dz$ (see Section 2.4), integrating along the lines of sight 
gives the projected two-dimensional distribution (Peebles 1980): 
\begin{equation}
\w = A_{\omega}\theta^{1-\gamma},
\label{eq:ampl}
\end{equation}
where $\theta$ is in radians and $A_{\omega}$ is given by (e.g. Phillipps \et
1978):
$$ A_{\omega} = C r_{0}(0)^{\gamma} \int_0^\infty D_{\theta}^{1-\gamma}(z) g^{-1}(z)
(1+z)^{-(\epsilon + 3)} \left( {dN \over dz} \right)^2 dz
      \left[ \int_0^\infty  \left( {dN \over dz} \right) dz \right]^{-2}$$
where $D_{\theta}(z)$ is the angular diameter distance, 
g(z) is the scale factor multiplied by the element of comoving distance
$d\omega/dz$:
\begin{equation}
g(z) = \frac{c}{H_{o}} ( (1+z)^{2}(1+\Omega_{o}z)^{\frac{1}{2}})^{-1}
\end{equation}
and C is a constant involving only numerical factors:
\begin{equation}
C = \sqrt{\pi}\frac{\Gamma[(\gamma-1)/2]}{\Gamma(\gamma/2)}.
\end{equation}

Note that the above amplitude depends (strongly) on the shape of the redshift 
distribution, $dN/dz$ (i.e. effectively, the ``width'' squared and the
median redshift), but not on the 
overall normalization. To perform this inversion we have assumed that 
$\w$ is a perfect power law, i.e. on all scales, and that the slope,
$\gamma$ does not change with epoch, consistent with our observations.

In principle, knowing $N(m,z)$ at all magnitudes, one could fit all of the data
in Figure ~5 and determine $r_0 (z)$ at all z.   
However, there are at least two effects to consider:
the rest-frame waveband of the observed R-band shifts with redshift,
and different populations of galaxies may evolve quite differently in
luminosity (see e.g. Lilly \et 1995c). Together these imply that the
morphological composition of the sample changes with depth.
As clustering characteristics can vary with galaxy type (e.g. Loveday
\et 1995), any apparent evolution in clustering may reflect a change
in the galaxy population rather than the evolution of density perturbations
in the Universe.
Therefore, in producing an evolving correlation function, the evolution of 
large scale structure and the evolution of the galaxy population 
are likely to be inextricably linked. We are thus determining an ``effective''
$\epsilon$ that reflects both effects.

In order to reduce our measurement of $\w$ at $R=23.5$ to a single useful
number, we seek simply to
estimate the value of $r_{o}(z)$ at some typical redshift of the sample.
Determining the best value of $r_{0}(z)$ at high $z$, our procedure 
was to find all combinations of 
$r_{0}(0)$ and $\epsilon$ that are consistent with the correlation amplitude 
for our galaxy sample
at $19 < R < 23.5$. It is found that these different ``consistent'' 
$r_{0}(z)$'s all intersect at a redshift of z = 0.48 at a value 
$r_{0}(0.48) = 1.86 \h$ Mpc for $q_0 = 0.5$ and $r_{0}(0.48) = 2.16 \h$ 
Mpc for a low-density Universe,$q_0 = 0.05$. 
We note that an intersection of $z\sim 0.48$ is convenient since, at this
redshift, the observed R-band is approximately equivalent to the rest-frame
B-band, allowing a ready comparison between our faint sample and local 
B-selected samples.
It should be noted that, the intersection is somewhat less than 
the median redshift of the sample $\langle z \rangle \sim 0.56$ due to the
higher weighting of lower-redshift objects in 
Equation \ref{eq:ampl}.

\subsection{Errors in the correlation length}
\label{ssec:errcl}

There are three contributions to the error in the correlation length. 
The first is
the error in the amplitude of $\omega(\theta=1^\circ)$ at $R=23.5$ that we
are trying to fit. 
This error is determined by fitting each of the 300 bootstrap realizations
to a slope of $\delta=-0.8$ and taking the standard deviation of the
resulting amplitude distribution.
We have found $\omega(\theta =1^\circ)=0.001386 \pm
0.00024$ which corresponds to a fractional error in $\rnotz$ of 18\%. 

The second contribution to the error in the correlation length is from the 
galaxy fraction. An estimate of the error in the galaxy fraction is 
$10\%$ (from the relative numbers of spectroscopically identified stars 
and galaxies in the 22hr field of the CFRS),
corresponding to a fractional error in $\rnotz$ of $14\%$.
This was determined by varying the amplitude $\omega(\theta=1^\circ)$ by 
10\% and examining the range in $\rnot$ for $\epsilon=0$, and
hence $\rnotz$, that this implies.

The third contribution to the error is from the redshift distribution,
$N(z)$. A total of 730 galaxies from the CFRS were used in the R-band
$N(z)$ determination, which was then put into Limber's equation, with
$\rnot$ and $\epsilon$ as free parameters, to determine $\rnotz$.
We have investigated the statistical error introduced by errors in $N(z)$ by 
resampling the galaxies that went into the $N(z)$ 
determination and comparing the fit to the correlation amplitude for
a given $\rnot$ and $\epsilon$. We have performed 20 resamplings
and find that the contribution to the fractional error in
$\rnot$ due to purely statistical uncertainties in the redshift distribution 
is approximately 3\%, much smaller than the above two sources of error.

Adding these three contributions in quadrature gives $\delta\rnotz = \pm 0.43$.
It is noted that the dominant source of error in $\rnotz$ is in the
amplitude of $\w$.
This error in $\rnotz$ corresponds to an uncertainty in $\epsilon$ of 
$\pm 1.1$.

An additional, systematic, effect to note is that redshifts were 
unobtainable for 15\% of the objects in the CFRS. Crampton \et (1995) have
argued that at least half of the unidentified 15\% must be distributed like
the CFRS sample as a whole. If the other half of the unidentified 15\%
were located at high redshifts, they would act to dilute the predicted
correlation. Hence, in comparing to observations, a higher predicted 
correlation length must be chosen to account for the extra dilution. We
find that such an effect would increase the deduced correlation
length by approximately 10\%.

\section {Discussion}
The deduced $r_{0}(0.48) = 1.86 \h$ Mpc
can be produced by the following combinations:
$\epsilon = -1.2$, $r_{0}(0) = 2.75
\h$Mpc; $\epsilon = 0$, $r_{0}(0) = 3.60 \h$ Mpc; $\epsilon = 1$, 
$r_{0}(0) = 4.45 \h$ Mpc. These possibilities are shown in Figure~5.
As discussed above, although the high $\epsilon$ curves nominally provide a 
better fit
to the data over a wide magnitude range, the fact that the galaxy population 
may possibly be changing 
with depth (i.e. redshift) means that a firm conclusion to this effect can 
not be drawn.

The range of possible $r_{0}(0)$ and $\epsilon$ combinations is
illustrated in Figure~6, 
which shows the three representative models which are constrained to go
through
the $r_{0}(z=0.48) = 1.86 \h$ Mpc. These are compared to several local surveys
tabulated in Table 2.
The CfA survey has a limiting magnitude of $B \sim 14.5$ and contains 1840 
galaxies in the North zone, of which 1230 were used in the correlation
analysis (Davis and Peebles 1983). 
The APM/Stromlo survey is complete to $b_{J} = 17.15$ and contains 1769 
galaxies (Loveday \et 1992).
The ``Durham'' survey represents three surveys with 676 redshifts to a 
limiting depth of $b_{J} \sim 16.8$ (Hale-Sutton \et 1989).
The KOS survey consists of redshifts for 164 field galaxies brighter than
$J = 15.0$ in the north and south galactic caps.
The IRAS redshift survey represents results from both the $S_{60} \geq 0.6$ Jy 
(Saunders \et 1992) and $S_{60} \geq 1.2$ Jy (Fisher \et 1994) catalogs.

As a first statement, we can conclude from Figure~6 that our faint 
galaxies at $z =0.48$ can ``easily" evolve into the
local population (with correlation length between $r_{0}(0) =$ 4.0 and 
5.5 h$^{-1}$ Mpc) if the clustering evoluton is given by $0 < \epsilon < 2$, 
i.e. with mild to rapid evolution.
In other words, the galaxies observed
down to $R=23.5$ cluster like ``normal'' galaxies.
At a more detailed level, evolution to a CfA-like population requires 
$\epsilon \sim +2$ and evolution to an IRAS-like population requires 
$\epsilon \sim 0$. It is important to note that clustering evolution with 
$\epsilon \sim -1.2$
(i.e. fixed in comoving space) would result in a local population that is 
considerably less clustered ($r_{0}(0) \sim 3h^{-1}$ Mpc) than {\it any} of 
the local samples plotted in Figure ~6.  While low values 
of $r_{0}(0)$ have been measured for local dwarfs (e.g. Santiago
and da Costa, 1990), these have yet to be confirmed (e.g. Thuan \et 1991).
These conclusions generally remain intact when a low-$\Omega$ Universe is
considered. Changing $\Omega$ from 1.0 to 0.1 affects the line element and
the angular diameter distance thereby increasing the correlations at
large distances and hence reducing the implied clustering evolution to local 
samples. In this case, the range in local correlation length, 
$\rnot = 4.0-5.5 \h$ Mpc is satisfied by $\epsilon \sim -0.5$ to $1.0$.

Can our knowledge of the galaxy population be used to distinguish between these
possibilities?  Two results from the analysis of the luminosity function of 
the I-band-selected Canada-France redshift survey (Lilly \et 1995c), 
selected to have $17 < I < 22 $ and thus broadly equivalent to the present 
R-band selected sample, are relevant.  First, the luminosity function of
the redder galaxies (redder than a typical local Sbc galaxy) shows no change 
back to $z \sim 1$, 
suggesting that these galaxies have been relatively stable over the time 
interval relevant to the 
present study.  These galaxies are presumably the same galaxies that appear in
local B-selected samples such as the CfA.  On the other hand, the luminosity 
function of the
bluer galaxies shows substantial evolution. At $z =0.48$ there are about 
three times more 
blue galaxies with luminosities comparable to present-day L* than are found 
at lower redshifts.
These galaxies, which dominate the galaxy population at $z \sim 0.6$, are 
responsible for the steep number counts of galaxies in the B-band. Their 
nature and the identity of their local descendents (and thus their
expected $r_{0}(0)$) are not yet determined.  
Important clues as to the nature of these galaxies are given by
Le F\`evre \et (1996) who have determined correlation lengths for the
blue and red populations separately.
At $z>0.5$ blue and red galaxies are found to cluster similarly while at
lower redshifts red galaxies cluster more strongly than blue galaxies.
This indicates that environment may be playing a role in the relationship
between blue and red galaxies as the Universe expands.

Returning to Figure ~6, we now comment on the much fainter Efstathiou \et (1991)
and Brainerd \et (1995) results.
We note that they lie well below the predicted curves for any of 
the models in Figure ~6
which were constructed from the expected $N(z)$ and the observed $\w$ at 
$R < 23.5$. That is, the expected levelling off in the amplitude of the 
correlation function with depth (due to the fact that $N(z)$ is expected
to cease changing with depth because of the effects of volume constriction
at high redshifts) is not observed. This may reflect 
the fact, noted above, that the $N(m,z)$ analysis (derived from the 
Canada-France Redshift Survey sample at 
$I < 22$) fails to reproduce the number counts at the faintest magnitudes 
(see Lilly \et 1995c), presumably because of evolution in the galaxy sample 
at these faintest magnitudes relative to that seen in the CFRS at $I \sim 22$
or in our sample at $R<23.5$.

\section{Summary}

The angular correlation function has been determined to a limiting magnitude
of R = 23.5 for a galaxy catalog containing 13000 objects in a region 4
deg$^2$ using high quality 
sub-arcsecond seeing CFHT images. The main results are:

\begin{itemize}
\item
The amplitude of the correlation function at arcmin-scale separations 
is in accord with most other recent determinations.  

\item No evidence for a 
significant decrease in the slope of $\w$ away from $\delta \sim 0.8$ is seen.

\item
Using the $N(z)$ information from the I-band selected Canada-France Redshift 
Survey to predict an N(z) for this sample, 
we estimate the
correlation length at $z = 0.48$ to be $r_{0}(0.48) = 1.86 \pm 0.43 \h$ Mpc 
(for $q_0 = 0.5$) and $r_{0}(0.48) = 2.16 \pm 0.49 \h$ Mpc 
(for $q_0 = 0.05$).  
\item
This is consistent with
local samples of normal, optically
selected galaxies if clustering has grown in the Universe since 
$z \sim 0.6$.
Evolution to CfA-like samples with  $r_{0}(0) \sim 5.5 \h$ Mpc would 
require clustering evolution
with $\epsilon \sim +2$, stronger than that seen in CDM-like hierarchical 
models, while evolution
to IRAS-like samples with $r_{0}(0) \sim 4 \h$ Mpc would require 
evolution with $\epsilon \sim 0$, which can be obtained with clustering of 
fixed physical size. Clustering that is fixed in comoving space, i.e. 
$\epsilon \sim -1.2$, would result in a local sample with very weak 
clustering, $r_{0}(0) \sim 3 \h$ Mpc.
\end{itemize}

\section{Acknowledgements}
This investigation benefitted greatly from the N(z) information that was 
derived
from the Canada-France Redshift Survey and SJL thanks his collaborators on 
that project,
Olivier Le F\`evre, Francois Hammer, David Crampton and Laurence Tresse,
for their indirect contribution to the present work.
We also thank Arif Babul, Ray Carlberg, 
Nick Kaiser (who also helped take the data), 
Omar Lopez-Cruz, Chris Pritchet, Dave Schade, Chuck Shepherd and
David Woods for stimulating conversations, Frank Valdes for assistance
with FOCAS, and Howard Yee for the use
of PPP. 
We gratefully acknowledge the two detailed reports on this paper made by
the referee, Dick Fong.
In making these observations we were ably assisted
by the Telescope Operators at CFHT,
Ken Barton, John Hamilton and Norman Purves.
This research was funded by NSERC of Canada and by the University of Toronto.

\newpage
\section{References}

\apjs{Bahcall, J. and Soneira 1980}{44}{73}
\mn{Brainerd, T.G., Smail, I. and Mould, J. 1995}{275}{781}
\mn{Couch, W.J., Jurcevic, J.S. and Boyle, B.J. 1993}{260}{241}
\apj{Crampton, D., Le F\`evre, O., Lilly, S. and Hammer, F., 1995}{455}{96}
\apj{Davis, M., Efstathiou, G., Frenk, C., and White, S.D.M., 1985}{292}{371}
\apj{Davis, M. and Peebles, P.J.E. 1983}{267}{465}
\stat{Efron, B. and Tibshirani, R. 1986}{1}{54}
\apj{Efstathiou, G., Bernstein, G., Katz, N., Tyson, J.A. and Guhathakurta, P.}{380}{L47}
\mn{Fisher, K.B., Davis, M., Strauss, M.A., Yahil, A. and Huchra, J. 1994}{266}{50}
\apj{Groth, E.J. and Peebles, P.J.E. 1977}{217}{385}
\mn{Hale-Sutton, D., Fong, R., Metcalfe, N. and Shanks, T. 1989}{237}{569}
\mn{Hewett, P. C. 1982}{201}{867}
\refbook{Hudon, J.D. 1995, PhD thesis, Univ. Toronto}
\aa{Infante, L. 1994}{282}{353}
\apj{Infante, L. and Pritchet, C.J. 1995}{439}{565}
\aj{Kirschner, R.P., Oemler, A. and Schechter, P.L. 1978}{83}{1549}
\aj{Landolt, A.U. 1983}{88}{439}
\apj{Le F\`evre, O., Crampton. D., Lilly, S.J., Hammer, F. and Tresse, L. 
1995}{455}{60}
\apjpr{Le F\`evre, O., Hudon, D., Lilly, S.J., Crampton, D., Hammer, F
and Tresse, L. 1996}
\apj{Lilly, S.J. 1993}{411}{501}
\apj{Lilly, S.J., Cowie, L.L. and Gardner, J.P. 1991}{369}{79}
\apj{Lilly, S.J., Le F\`evre, O., Crampton. D., Hammer, F. and Tresse, L. 
1995a}{455}{50}
\apj{Lilly, S.J., Hammer, F., Le F\`evre, O., and Crampton, D., 1995b}{455}{75}
\apj{Lilly, S.J., Tresse, L., Hammer, F., Crampton. D. and Le F\`evre, O. 1995c}{455}{108}
\mn{Ling, E.N., Frenk, C.S. and Barrow, J.D. 1986}{223}{21p}
\apj{Loveday, J., Peterson, B.A., Efstathiou G., and Maddox, S.J. 1992}{390}{338}
\apj{Loveday, J., Maddox, S.J., Efstathiou, G. and Peterson, B.A. 1995}{442}{457}
\mn{Maddox, S.J., Efstathiou, G., Sutherland, W.J. and Loveday, J. 1990}{242}{43p}
\mn{Metcalfe, N., Shanks T., Fong, R., Jones, L. R. 1991}{249}{498}
\refbook{Peebles, P.J.E. 1980, ``The Large Scale Structure of the Universe'',
(Princeton), p. 185.}
\mn{Phillipps, S., Fong, R., Ellis, R.S., Fall, S.M. and MacGillivray, H.T. 1978}{182}{673}
\aj{Pritchet, C.J. and Infante, L. 1986}{91}{1}
\mn{Roche, N., Shanks, T., Metcalfe, N. and Fong, R. 1993}{263}{360}
\apj{Santiago, B.X. and da Costa, L.N. 1990}{362}{386}
\mn{Saunders, W., Rowan-Robinson, M. and Lawrence, A. 1992}{258}{134}
\aa{Sharp, N.A. 1979}{74}{308}
\mn{Stevenson, P.R.F., Shanks, T., Fong, R. and MacGillivray, H.T., 1985}{213}{953}
\apj{Thuan, T.X., Alimi, J.-M., Gott, J.R. and Schneider, S.E. 1991}{370}{25}
\refbook{Valdes, F. 1982, ``FOCAS User's Manual'', (Tucson:NOAO)}
\pasp{Yee, H.K.C. 1991}{103}{396}

\newpage
\vskip 1.5cm
\noindent{\bf FIGURES}
\vskip 0.5cm

\noindent{\bf Figure~1:}
Galaxy and star counts for our data compared to the recent compilation of 
results in Metcalfe \et (1991) shown as the dotted lines.
Poisson errors are too small to be seen.
Also shown are the stars determined from our star-galaxy separator for 
magnitudes $R < 22.0$ (starred symbols) and 
the predictions from the galactic model of Bahcall and Soneira (x's).

\vskip 0.25cm
\noindent{\bf Figure~2:}
Results for our star-galaxy separator for four different fields. Stars
occupy the horizontal sequence which is well-defined until $R = 22.0$.
This was used to estimate the stellar contamination for $R \leq 22.0$.

\vskip 0.25cm
\noindent{\bf Figure~3:}
The N(z) distribution inferred from the I-band selected CFRS for magnitude 
bins (of constant width 4.5 mag) above
$R = 18.0, 18.5, 19.0$, etc., up to $R=25.0$. 
The distribution used in the present work $R=19.0 - 23.5$ is highlighted.

\vskip 0.25cm
\noindent{\bf Figure~4:}
The angular correlation function from the counts-in-cells method for three 
magnitude ranges. Both the global (filled circles) and local (open circles)
averaging methods are shown. 
The data is uncorrected for the integral constraint, any spurious field to field variations
or for stellar contamination.
The dotted line represents a slope of $\delta = -0.8$.

\vskip 0.25cm
\noindent{\bf Figure~5:}
The observed correlation amplitudes scaled to $\theta = 1^{\circ}$ for
this work (filled circles) along with several recent studies. 
A slope of $\delta = -0.8$ has been used for all points and stellar
contamination has been corrected for.
Three evolutionary models described in the text which reproduce our
$\w$ at $R = 23.5$ are also shown
These have combinations of
[$r_{0}(0) , \epsilon$] of [4.45,+1] (steepest), [3.6,0] (middle)
and [2.75,-1.2] (shallowest). See text for discussion.

\vskip 0.25cm
\noindent{\bf Figure~6:}
The correlation length for our sample at $z = 0.6$ ($q_{0} = 0.5$) compared 
to local results,
with four values of the evolutionary parameter, $\epsilon = -1.2,0,1,2$
(shallowest to steepest). Setting $q_{0} = 0.05$ gives the open circle.

\newpage
\begin{table}
\caption{Correlation function parameters}
\begin{tabular}{ccccccc}
\hline
\hline
mag. interval & $N_{obj}$ & $N_{galaxies}$ & $N_{gal} deg^{-2}$ & $f_{g}$ & 
$A_{\omega}^{\delta=0.8}$ & 
$log A_{\omega}^{corr}$ \\
\hline
$19.0 \leq R \leq 22.5$ & 6209  &  4153  &  14726 & 0.67 & 0.00128 $\pm$ 0.00017 & -2.54 $\pm$ 0.06\\
$19.0 \leq R \leq 23.0$ & 8856  &  6323  &  22421 & 0.71 & 0.00105 $\pm$ 0.00011 & -2.68 $\pm$ 0.05\\
$19.0 \leq R \leq 23.5$ & 12757 &  9710  & 34432 & 0.76 & 0.00080 $\pm$ 0.00014 & -2.86 $\pm$ 0.08\\
\hline
\end{tabular}
\end{table}

\begin{table}
\caption{Local Correlation Results}
\begin{tabular}{cccccc}
\hline
\hline
Survey & $z_{med}$ & $N_{galaxies}$ & $r_{0}(0)$ & $r_{0}(z_{med})$&References\\
       &           &                &(h$^{-1}$ Mpc)&(h$^{-1}$ Mpc)  &        \\
\hline
CfA    & 2500 km s$^{-1}$   & 1230 &  5.4   &   5.3 &Davis and Peebles (1983)\\
       & z = 0.0083         &      &        &         &                      \\
APM/Stromlo & 15,200 km s$^{-1}$ & 1769 &  5.7   &   5.2   & Loveday \et (1992) \\
       & z = 0.05           &      &        &         &                      \\
Durham & $\sim 13,000$ km s$^{-1}$ & 676 & 4.5 & 4.2  &Hale-Sutton \et (1989)\\
       & z = 0.043           &      &        &         &                     \\
KOS    & $\sim 8000$ km s$^{-1}$& 164 & 4.0 &    3.8   & Kirschner \et (1978)\\
       & z = 0.027           &      &        &         &                     \\
IRAS (0.6Jy)& $\sim 5000$ km s$^{-1}$ & 9080 & 3.8 & 3.7 &Saunders \et (1992)\\
(QIGC) & z = 0.017           &      &        &         &                     \\
IRAS (1.2Jy)& $\sim 5000$ km s$^{-1}$ & 5313 & 3.8 & 3.7 & Fisher \et (1994)\\
       & z = 0.017           &      &        &         &                   \\
\hline
\end{tabular}
\end{table}

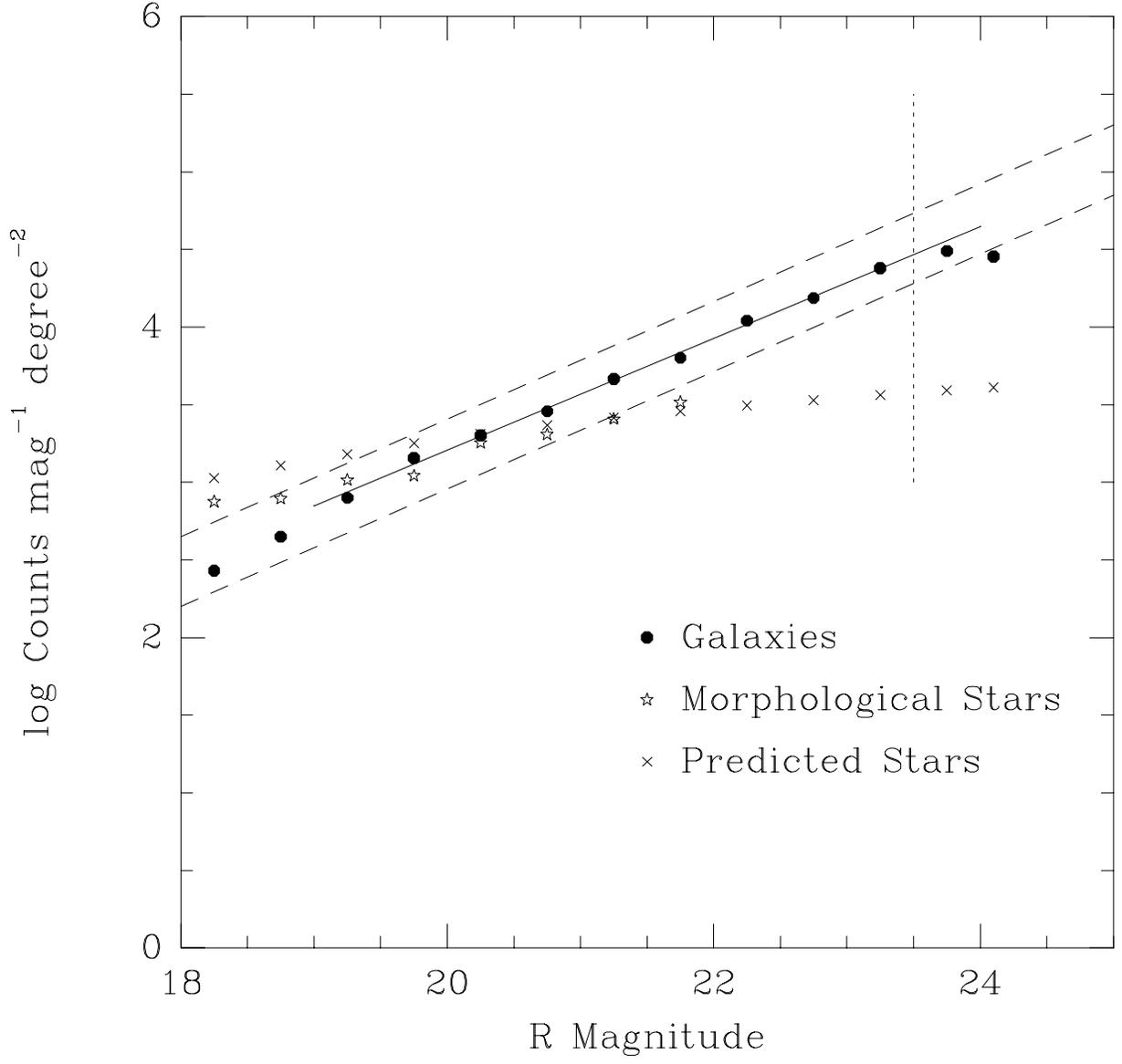
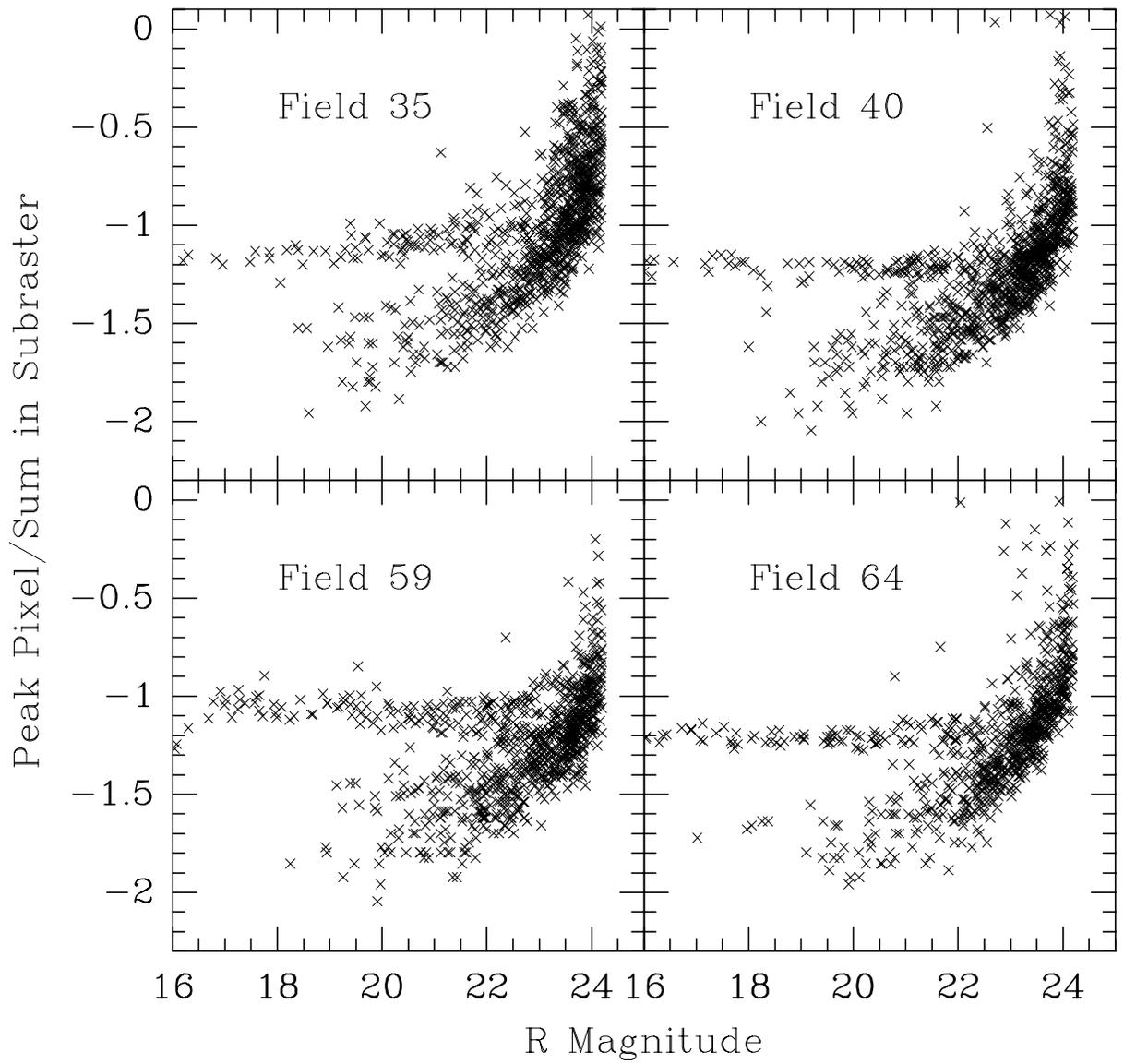
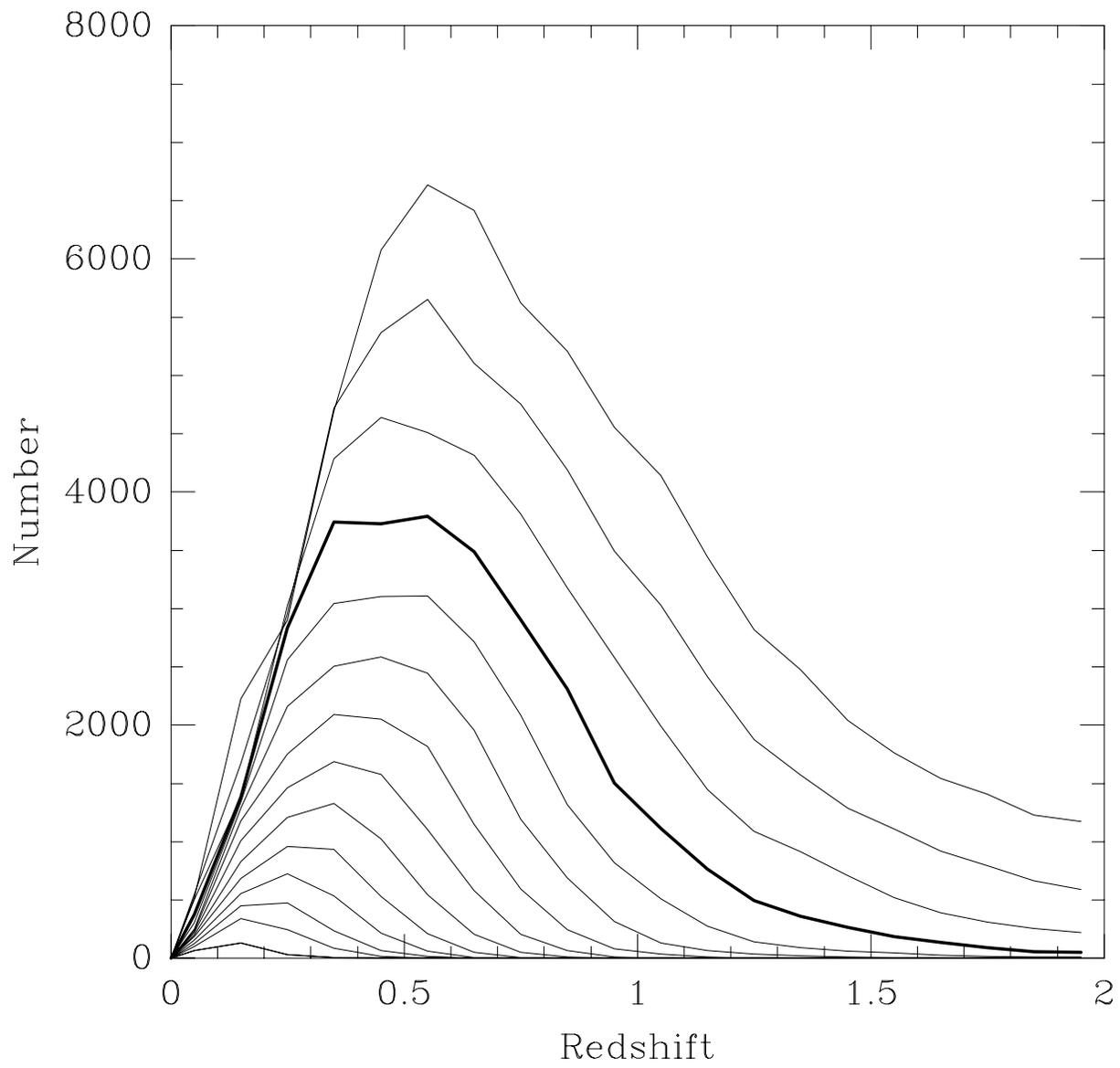
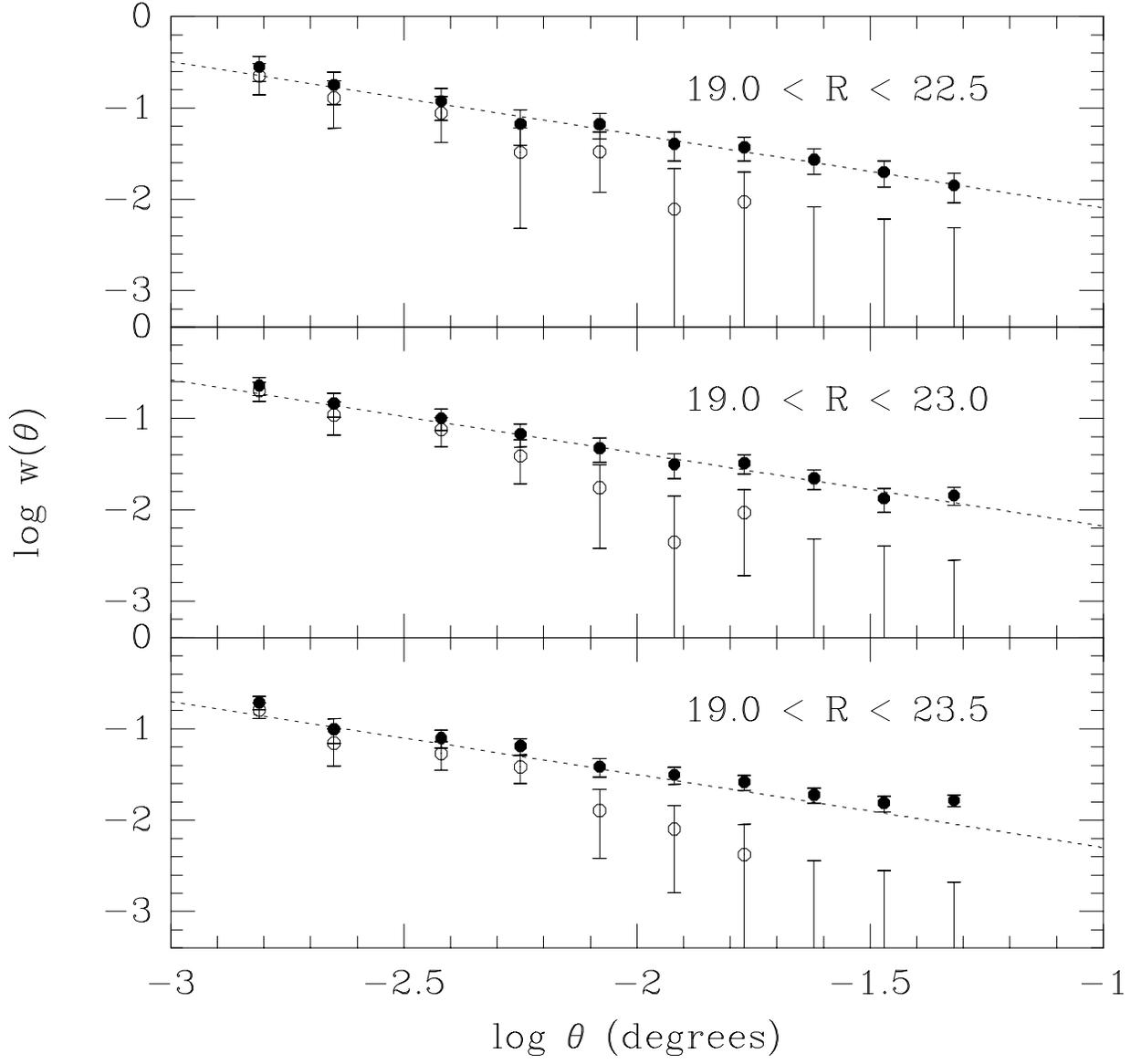
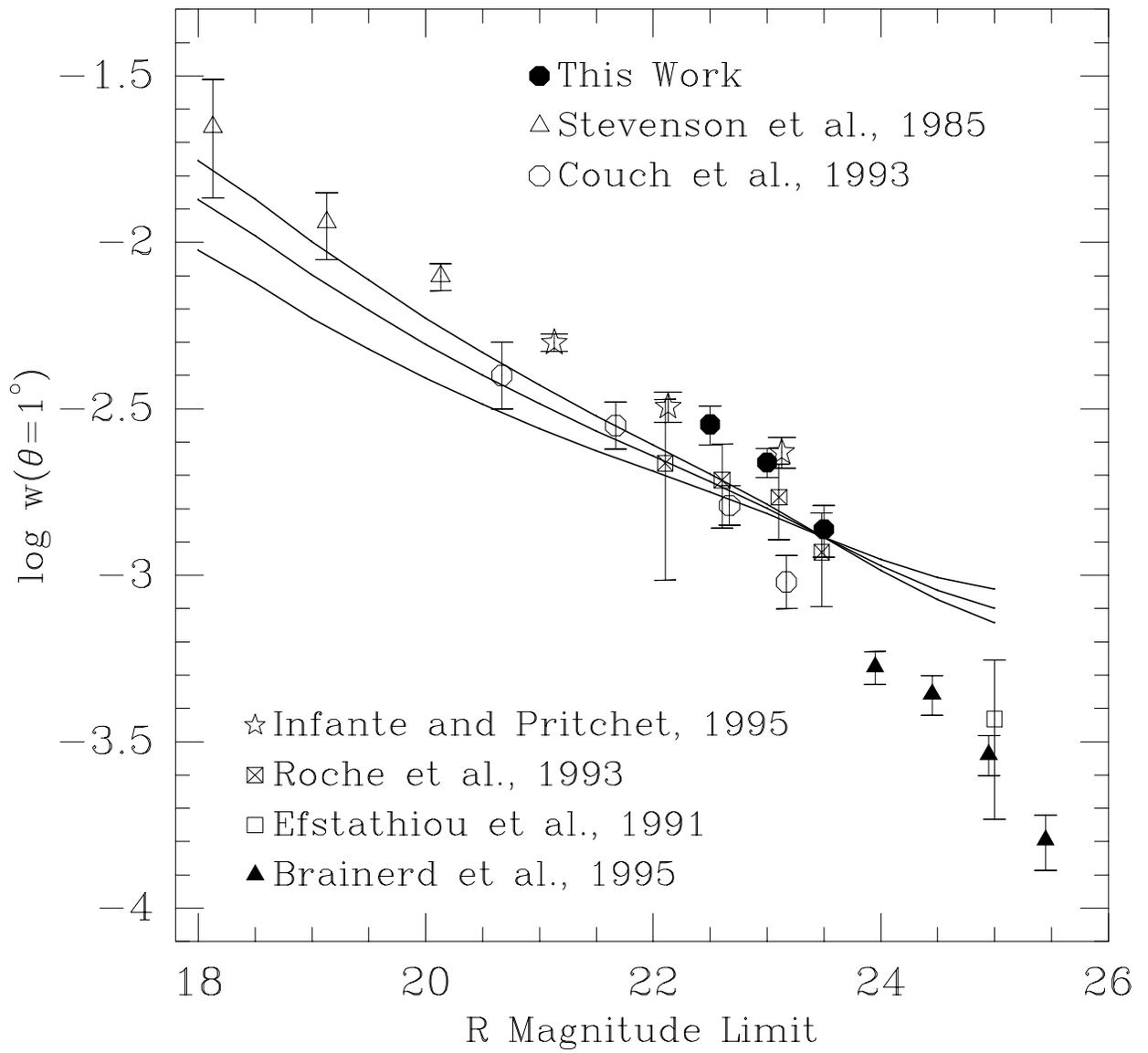
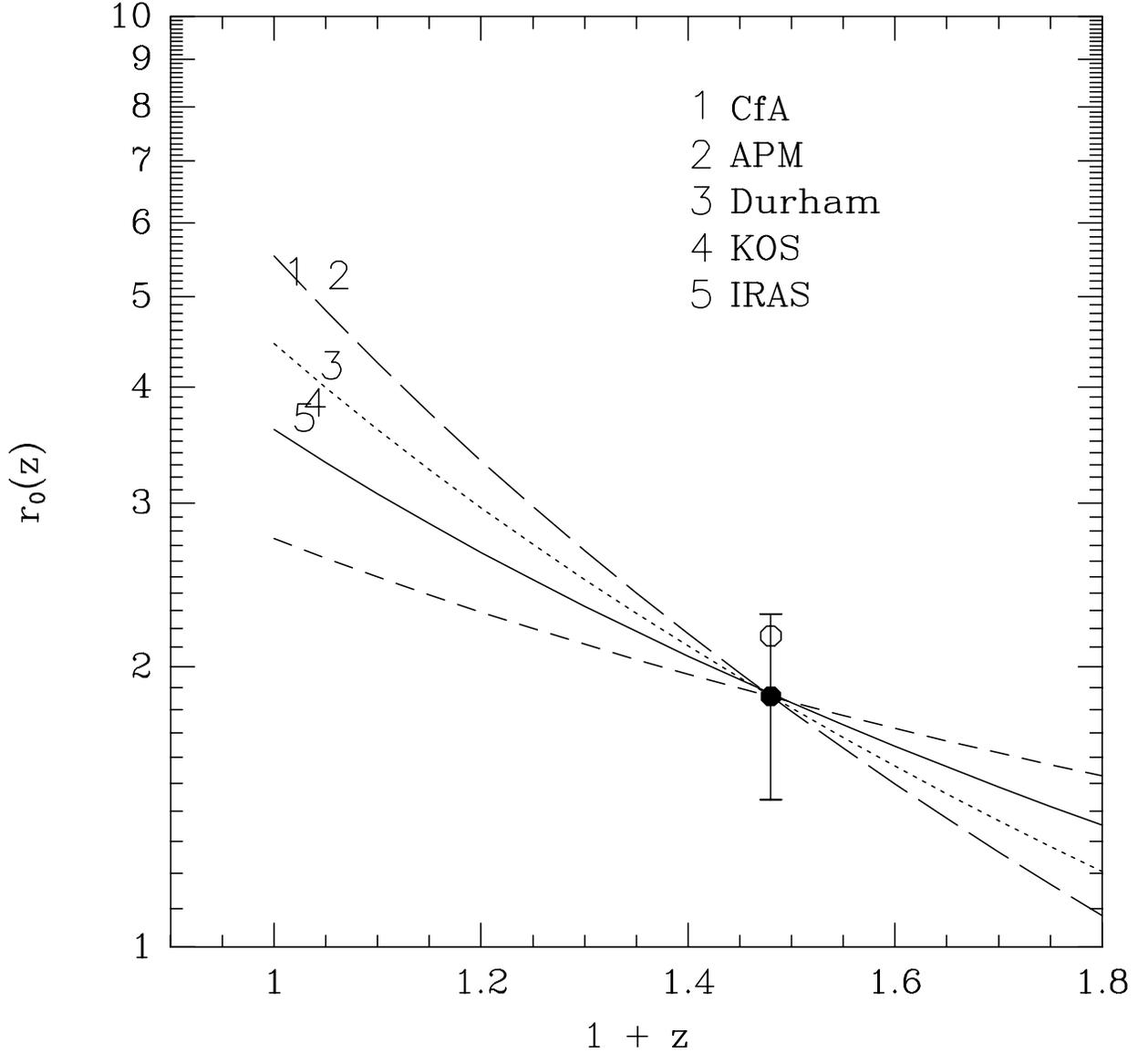
\end{document}